\documentclass[twoside,leqno,twocolumn]{article}

\usepackage[letterpaper]{geometry}

\usepackage{ltexpprt}
\usepackage{hyperref}
\usepackage{algorithmic}
\usepackage{graphicx}
\usepackage{textcomp}
\usepackage{xcolor}
\usepackage{listings}
\usepackage{tikz}
\usepackage{cite}

\begin{document}

\newcommand\relatedversion{}
\renewcommand\relatedversion{\thanks{The full version of the paper can be accessed at \protect\url{https://arxiv.org/abs/1902.09310}}} 

\title{\Large A Novel Feature Representation for Malware Classification}
\author{John Musgrave \thanks{Department of Computer Science, University of Cincinnati}
\and Temesguen Messay-Kebede \thanks{Air Force Research Lab, Wright-Patt Air Force Base}
\and David Kapp \thanks{Air Force Research Lab, Wright-Patt Air Force Base}
\and Anca Ralescu \thanks{Department of Computer Science, University of Cincinnati}}

\date{}

\maketitle


\fancyfoot[R]{\scriptsize{Copyright \textcopyright\ 2022 by SIAM\\
Unauthorized reproduction of this article is prohibited}}





\begin{abstract} \small\baselineskip=9pt In this study we present a novel representation for features of malicious programs.  This representation is based on hashes of data dependency graphs, which are directly tied to both the structure and operational semantics of a program.  We present a comparison with existing term frequency representations and show an increase in accuracy and robustness.  Existing methods of deep learning are often based on $tf-idf$ feature representations, and a more robust feature representation enables better classification and pattern recognition.\end{abstract}

\section{Introduction}
We present a method for constructing a novel feature representation and compare it to a $tf-idf$ representation on the same data.

\subsection{Related Work}

Machine learning techniques have been applied in many contexts to successfully identify malicious programs based on a variety of features. Many classification methods have been used for supervised learning including deep neural networks and support vector machines. Several datasets have been collected with various kinds of features, including assembly instructions, n-gram sequences of instructions and system calls, and program metadata \cite{souri2018state}, \cite{rawashdeh2021single}, \cite{kebede2017classification}, \cite{djaneye2019static}, \cite{chandrasekaran2020context}.

The focus of many studies applying machine learning techniques to malware analysis is the task of classification for the purposes of identifying unknown programming errors. Zhou et al. used a graph neural network (GNN) model to classify various types of C functions in order to determine semantic errors in their abstract syntax tree (AST) and program flow. The model was trained on a dataset of functions which were drawn from several executable binaries including the Linux kernel. Wang et al. developed a synthetic dataset of 3 million Python programs with class labels, and extracted function call graphs from AST graphs generated from tokenization to be used for training with a novel GNN design. Park et al. have used sequence modeling to identify potential optimizations in programs based on an intermediate representation. A program flow graph was extracted from this intermediate program representation and used in the sequence predictions \cite{zhou2019devign}, \cite{wang2020learning}, \cite{park2012using}.

Several studies have used control flow graphs as features in datasets used for malware classification tasks. Bruschi et al. have extracted control flow graphs from malware for the purposes of classification through comparing the graphs for isomorphism. Cesare et al. have presented several studies on the uses of control flow graphs in the classification of malware with efficient results \cite{bruschi2006detecting}, \cite{cesare2010fast}, \cite{cesare2013control}, \cite{cesare2010classification}.

In a previous study we have collected graph features of programs and performed a quantitative analysis of the structural properties of the program networks \cite{musgrave2022networks}.

Hashing of features has been performed in several studies applying machine learning to malware analysis. The focus of the hashing is often the semantics of a function in a high level language. Jang et al. successfully used a hash function on features of binary n-gram sequences to represent malicious programs. These were compared for similarity by their Jaccard index. The focus of their work was an approach from unsupervised learning, and an analysis of the clusters of the hashes obtained. They used a co-clustering approach to demonstrate feature correlation, and also implemented $k$-Nearest Neighbors classification with precision and recall above 90\%. Their features focused primarily on binary stings, but can be extended by the development of a custom hashing function. Liang et al. applied partial order preserving hashing via Gödel hashes to obtain an increase on existing benchmarks for program flow analysis. LeDoux et al. represented a program as a graph of atomic functions obtained from a semantic hashing approach. The similarity of the hashes represented a semantic similarity in the functions. However, basic block features were overlooked, and whether the function semantics are correlated across abstraction layers is still an open question. In a similar manner, Alrabaee et al. have used a $tf-idf$ representation along with other methods, Hidden Markov Models and graph kernels, to obtain a graph of function semantic representations for a program. This was accomplished by constructing a Bayesian network for each of the features collected \cite{jang2011bitshred}, \cite{liang2014fast}, \cite{ledoux2013functracker}, \cite{alrabaee2018fossil}.

\subsection{Motivation}
Machine learning methods of malware analysis are widely claimed to represent features of operational semantics across abstraction layers by models and classification accuracy. The results and accuracy of classification techniques are dependent on the feature representations used in these datasets. Features are extracted from datasets collected at specific levels in an architectural hierarchy. Many useful features for classification can be extracted at multiple points in the architectural hierarchy as discussed in previous studies, e.g. instruction n-grams, sequences and patterns of bytecode or hex representations, as well as graphs, n-grams, and sequences of system API calls. The feature representation selected determines the granularity available and the degree to which classification accuracy is correlated with and representative of semantics. However, analysis of malicious programs presents several obstacles for an accurate classification of programs based on their operational semantics. Class labels are often coarse grained, with one label representing the class of an entire program, without a clear method to provide increased resolution for supervised models which are dependent upon labeled data.  Without an increased resolution of features that are descriptive of structure, explaining correlation between structure and semantic abstraction is very challenging.  The degree to which a program's component parts contribute to a class cannot be determined without increased feature resolution in a labeled dataset.  Document analysis methods are often based on a term frequency representation, where term co-occurrence can be measured.  But class labels for features in labeled datasets are often at the most coarse grained level of the binary as a whole.  A program level of resolution is too low to provide meaningful information about the relationship between a class label and a program's operational semantics across abstraction layers.  Therefore we view feature representations from two perspectives: as a description of program operational semantics, and the ability to describe structural properties. Structural properties enable the syntactic elements of a program to be interpreted. Semantic properties allow the correctness of a program to be verified across abstraction layers.

Several approaches exist to analyze a program based on its behavior, including static and dynamic analysis, or collecting execution traces $a-posteriori$. In the case of malware analysis, a formal specification for a program does not exist. A binary executable is the sole artifact for analysis. For this reason it is necessary to take a bottom up approach to the structural analysis, rather than collect artifacts of higher level language descriptions. Studies based on malware graph features often have a focus on methods of differentiation in control flow graphs through graph isomorphism.  While this is a useful feature for the classification of malicious programs, it is only a partial view of the program's structure and operational semantics.  The isomorphism of graphs and all subgraphs is often used to determine class equivalence, and this is a combinatorial operation.

\begin{figure}[t]
   \centering
    \hspace{1cm}
    \begin{lstlisting}
        mov    ecx, rbp - 44
        mov    eax, ecx
        and    eax, 400
        or     eax, 140
        or     ecx, 1
        cmp    rip + 170, 0
        cmovne ecx, eax
        mov    rbp - 44, ecx
        mov    rip + 180, 0
        jmp    0x100000000
    \end{lstlisting}
    
    \vspace{1cm}
    
      \begin{tikzpicture}[node distance={15mm}, thick, main/.style = {draw, circle}] 
    \node[main] (1) {$a_1$}; 
    \node[main] (2) [right of=1] {$a_2$}; 
    \node[main] (3) [below of=1] {$a_3$}; 
    \node[main] (4) [right of=3] {$a_4$}; 
    \draw[->] (2) -- (1); 
    \draw[->] (1) -- (3); 
    \draw[->] (3) -- (1); 
    \draw[->] (1) -- (4);
    \draw[->] (4) -- (1); 
    \end{tikzpicture}
    \[ A_{ddg} = \{ a_i \ | \ a_i \ \in \ A_{operand} \} \]
    \caption{Basic block segment of assembly instructions and its data dependency graph.  The data dependency graph shown is constructed from data movement instruction dependencies.  $mov$ instructions are the primary instructions with respect to term frequency.}
\end{figure}

\begin{figure*}[t]
   \centering
    \hspace{1cm}
    \begin{lstlisting}
                P1 = {0: 'd065c6962d38654c57e4ba3e3cf7e54c',
                      1: '60e7bbbe5c520d9011c0e21333912de5',
                      2: '45a5cb23f4ecba58dcec1e04f92390a7',
                      ...
                      233: 'f8bdeba55cbbc00a8413f528b729527e',
                      234: '11d4db31c81f5617e52dc0998de02846'}
    \end{lstlisting}
    \caption{A program represented as a set of hashes.  The hashes collected represent a pattern of isomorphically unique data dependency graphs.}
\end{figure*}

Among the most common representations are $tf-idf$ features and data flow analysis of functions in high level languages.  However, as we discuss in this paper, $tf-idf$ does not provide increased feature resolution of components below the class label of whole program level analysis.  Data flow analysis of functions provide a structural description, but this description does not persist across abstraction layers.  It is a description of the higher level abstractions, of which many exist, and a correlation to lower levels would still need to be shown.  This analysis is often performed after reverse engineering, and no specification exists in an adversarial use case.  In previous studies we have analyzed the structural properties of program networks collected from data dependency, control flow, and program dependency graphs.  In the absence of a formal specification, semantics are represented by patterns in structural properties of the program. Networks provide such a representation, and can be analyzed for their structure. Constructing networks from the lowest architectural level is a direct representation of the instructions being executed, and does not require translation across architectural layers \cite{musgrave2022networks}.

\subsection{Outline}
Section 2 covers the method and outlines our solutions.  Section 3 is an empirical evaluation of the methods discussed.  Section 4 discusses the significance and impact of the results.  Section 5 contains a summary and conclusion.

\section{Method and Solutions}
In this section we describe the data collection process and the method of constructing the feature representation.

\subsection{Data Collection}
Since we focus on an adversarial case for program execution, a dataset was composed by selecting samples of malicious and benign binaries for multiple platforms.  For our study, we focus on results obtained by using our representation to compare the two classes.  We have selected a small set of unknown malicious binaries with a large number of known benign binaries on the Windows operating system and $x86/64$ architecture.  In the case we are analyzing, an unknown malicious binary is presented to the system with unobserved behavior and the goal of obfuscation.  Additionally, the binary is the sole artifact available as no specification exists for verification prior to execution.  Benign program binaries for Windows were taken from the $System32$ directory, which contains programs that perform standard operating system functions.  We also briefly discuss benign samples taken from the Linux system directory $/usr/bin$.  Malicious samples were taken from the public malware repository $theZoo$ for both Windows and Linux malware \cite{thezoo}.

\subsection{Reverse Engineering}
Given a malicious binary, we reverse engineer the binary to obtain its $x86/64$ assembly representation.  This was done with the GNU $objdump$ utility.  The result of this step is a single document containing an assembly representation of the program.

\subsection{Segmentation}
Sequences of instructions are not determined by their linear placement within the document as in the case of natural language documents.  The sequences of instructions that are executed within the program are dependent upon the control structure of the program.  This is represented by the program's control flow graph (CFG).  Each node in this CFG graph is a basic block of contiguous assembly instructions.  For this reason we have segmented the program document into basic block segments.  Each segment is a basic block of contiguous instructions, and this is indicated by a jump instruction, or other control transition.

\subsection{Data Dependency Graph Extraction}
One data dependency graph artifact exists for each program segment.  We analyze the structure of dependency that exists between data operands.  Since the overwhelming prevalence of instructions are data movement, we focus on constructing data dependency graphs between $mov$ instructions.  This can be expanded to include all data operands, but the tail of the term frequency distribution is very thin, such that most of the variance is data movement.  By analyzing this structure of dependency, we can construct a graph artifact.  This process is shown in Figure 1.

\begin{figure*}[t]
    \centering
    \includegraphics[width=0.85\textwidth]{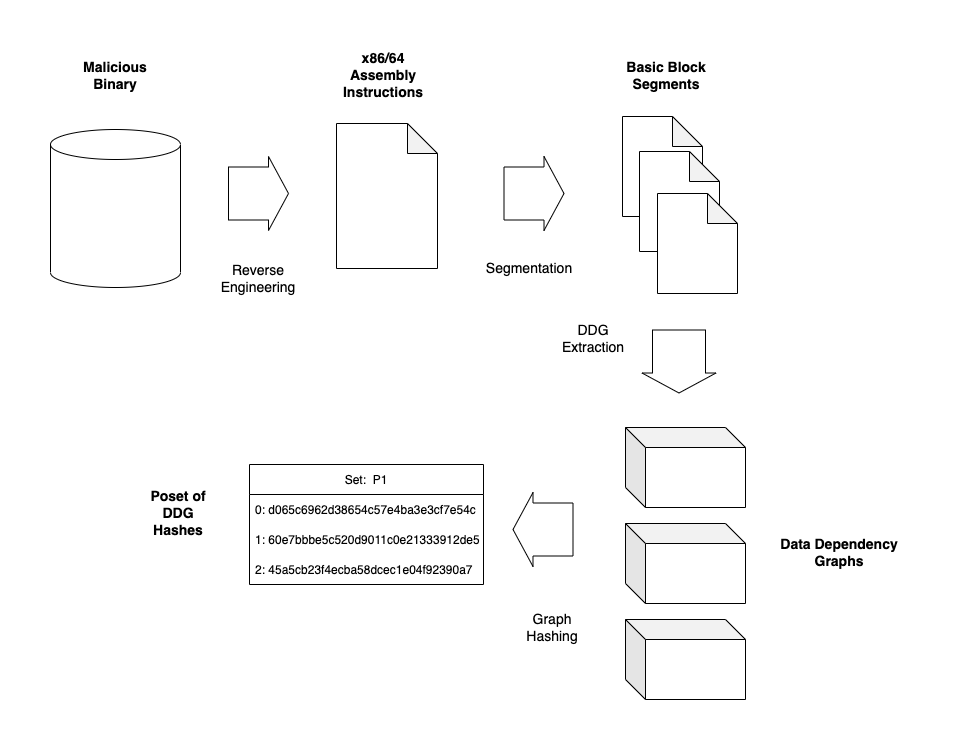}
    \caption{Feature extraction process.  Data dependency graphs are focused on patterns of data movement.  We use the Weisfeiler-Lehman graph hashing algorithm to compare graphs for isomorphism.}
\end{figure*}

\subsection{Graph Hashing}
Each graph was then hashed for uniqueness using the Weisfeiler-Lehman graph hashing algorithm.  This gives us a hash which represents a single graph, such that two isomorphic graphs will correspond to the same hash value \cite{shervashidze2011weisfeiler}. 

We are focused on capturing isomorphically unique graphs, and do not consider subgraphs or products.  For this reason, program segments generating duplicate hashes were not considered.  If a graph pattern exists in a program, this is sufficient to be used due to the transitivity of a program not being limited.  Therefore we do not analyze the frequency of graph usage at this point.

In this manner we can represent a program as a partial ordered set of hashes. Each hash corresponds to a basic block. Since we have segmented the program into basic block segments and have extracted data dependency graphs for each block, we can construct a set of hashes.  Since the dependency structure of a program's data is directly correlated to its operational semantics, a poset of DDG graph hashes is tied to both the structure and operation of a program.

\subsection{Set Partial Order}
The partial order of the set comes from the control flow graph structure of the program.  Since we have segmented the program into basic block segments, each of these segments can be considered an edge in the program's control flow graph.  An directed edge between nodes in the graph is obtained from a jump instruction from a node to a target.  Several utilities exist for recovering program control flow graphs, including $radare2$, which we have performed in other studies \cite{nar2019analysis}, \cite{musgrave2022networks}.

\subsection{Term Frequency Representation}
Term frequencies are a common way to extract feature vectors from documents. In order to construct a $tf-idf$ representation, a term dictionary must first be selected. Then, the frequency of each term in the document is measured. The resulting vector has one dimension per term, and the value is the frequency of the term in the document.  \cite{manning1999foundations}, \cite{shawe2004kernel}.

A dictionary was selected by collecting each opcode term from the $x86/64$ instruction set architecture, and then combining like operations via stemming. This yields a term dictionary of 32 unique terms.

In order to construct a $tf-idf$ representation, a term dictionary must be selected. Then, the frequency of each term in the document is measured. The resulting vector has one dimension per term, and the value is the frequency of the term in the document. It is also important to note that this $tf-idf$ vector represents a basic block segment of contiguous instructions. This will also provide an increase in accuracy to the $tf-idf$ representation over studies that were performed at a document level of granularity.  The result of the dictionary selection is a set of vectors, one per document, where each feature represents the term and each value is the term frequency.  The results of the term frequency analysis are heavily positively skewed towards data movement, such that the distribution tail is very thin.  Data movement instructions and their frequencies make up more of the distribution than the combination of other terms.

\section{Empirical Evaluation}
One question we may ask at the beginning of our analysis is:  what is the degree of similarity that a new program has to an existing program?  Let us first select a malicious sample from the dataset, one file from the $ZeusGameover\_Feb2014$.  A naive approach would compare across operating systems, and so we can compare this malware sample to the GNU/Linux \textbf{$ls$} program.  As discussed previously, normal operating system functions may be considered insecure by malicious actors depending on the context in which they operate.  So it is not obvious what normal operating system functions belong to the class of benign programs, if they contain potentially insecure functionality.  However, $ls$ is likely to only read information from the filesystem.  We expect these samples to not share many functional elements.

The total number of data dependency networks collected is 234 for $ls$ and 622 for $ZeusGameover\_Feb2014$ sample 1.  The set difference between the two sets will give us the degree to which the two programs are unique and differ from each other.  The number of data dependency graphs that are present in $ls$ that are not present in the $ZeusGameover\_Feb2014$ sample is 121.  The $ZeusGameover\_Feb2014$ sample \textbf{set difference} $ls$ has 509 unique data dependency graphs.

\subsection{What is the degree of overlap between two programs?}
We can measure the common data dependencies as the common functional patterns of data dependency between the two programs with the set intersection operation: \[ A \ \cap \ B \]

The intersection of $ls$ and $ZeusGameover\_Feb2014$ is 113.

The degree of overlap between two sets can be determined by the Jaccard coefficient, which is the ratio of cardinalities of the set \textbf{intersection} and \textbf{union} \cite{suppes1972axiomatic}, \cite{leskovec2020mining}.
\[ | A \ \cap \ B | \ /\ | A \ \cup \ B | \]

Since the set intersection is 113, we then calculate the union, which is 630.  The Jaccard coefficient is then 113/630, or 0.179.

\begin{figure}
    \centering
    \includegraphics[width=0.4\textwidth]{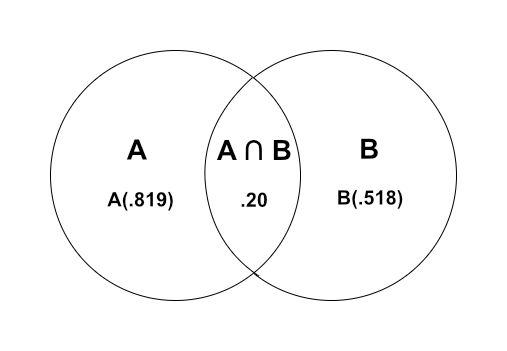}
    \caption{The overlap between malicious and benign samples is given by the Jaccard coefficient.  This means that a median value of 20\% of the structure of data movement within a program is shared between malicious and benign samples.  However, in the median example, 80\% of the functionality of malware is not shared with benign samples, and 50\% of benign functionality is not shared in the malicious sample, or described by the set intersection.}
\end{figure}

The dataset selection determines the results obtained from this representation to a large degree.  The degree of overlap that samples will have depends on both the malicious and benign samples selected.  However, this is one of the strengths of the representation as it offers increased feature resolution.  We have performed experiments with Linux targeted malware, and these malicious binaries do not share much with the benign programs on the same architecture.  But, without increased resolution, it is difficult to interpret these results.  For instance, we can repeat this process for the malicious Linux binary $Linux.Wirenet$ and benign samples collected from the $/usr/bin$ directory, which contains Unix system resources.  These samples share at most a Jaccard coefficient of 0.270 with the malware.  The minimum amount of overlap for these samples is 0.064, with a median of .204 Jaccard overlap.  This means that the malicious binary typically shares 20\% of its functionality with benign programs, and that these programs are also 80\% dissimilar from the malware.  Without an increase in feature resolution, it is challenging to know which properties are important to the class, as well as how much the classes overlap, or how to identify additional patterns within the class.  These tasks are critical in an adversarial use case.

If we expand our benign dataset of Windows programs, then we can compare the $ZeusGameover\_Feb2014$ sample to a larger class of functionality in known system programs.  So, we can add the benign class of programs from the $System32$ directory to our dataset for analysis.  We can then ask the question, what is the greatest degree of overlap between the malicious program and the benign class examples?  There is one sample that has a Jaccard coefficient of 1 with the Trojan malware.  Based on the structure of data dependency we can discover that $ZeusGameover\_Feb2014$ contains as a proper subset the system program \textbf{csrss.exe}.  This utility is the "Client/Server Runtime Subsystem" for Windows.  $csrss.exe$ is also used as a mechanism for Trojan malware to corrupt a system.  While not every Trojan malware uses this mechanism, it is often exploited by Trojan malware.  It is suspicious behavior for an unknown binary to include this program as a proper subset, as a legitimate user would likely have privileges to access the existing system utility based on their level of access.  This relationship between the programs was discovered from the analysis of the structure of their data dependency graphs, and by collecting hashes for each graph into sets.

\section{Significance and Impact}

\subsection{Discussion}
Why is this an effective representation? Our work focuses on the adversarial use case for program analysis. If program specifications were available for the program, then the method of verifying the semantic correctness of program operation would be a comparatively simple process. However, these specifications do not exist for malware because it is both novel and adversarial in nature. So a bottom up approach to feature construction must be taken. Methods of proving correctness of operational semantics exist at higher levels of abstraction, but it would need to be shown that these translate to the lower abstraction level of assembly instructions and data operands. This lower abstraction level is the primary representation available for a malicious binary, and proof or verification across abstractions requires further analysis.

This set of features allows for the resolution to be adjusted to obtain more or less information, depending on the goals of the analysis. It is not clear how a $tf-idf$ representation would provide increased resolution when given a distribution that is increasingly skewed towards one operation. The reason for this is that additional information describing the structural properties are lacking in the term frequency representation. The description of the structure provides an increase in resolution, and the structural properties described are that of data movement through the structure of their dependency. This provides additional information beyond the frequency of co-occurrence of the term. Structural properties are described by the representation of dependency, and operation is described by the properties of data movement.  By having a representation tied to the operands, we can describe the dependencies between the terms.  Therefore the data dependency graphs are directly descriptive of both the program's structure and operation.  The representation of dependency is able to provide additional structure to the body of the term frequency distribution, which is the goal of this study.

Security is largely contextual, and certain behaviors may be considered secure by a given actor in a given context, and insecure in the same context by another actor. For example, deleting a file is part of the necessary functioning of a filesystem, but is also a malicious task when a file is deleted improperly.  Not only is the execution of malicious binaries adversarial, it is also intentionally obfuscated.  So the goal of the adversarial party is to have the malicious properties have as much overlap with benign properties as possible.

The Weisfeiler-Lehman hashing method ensures that each isomorphically unique graph corresponds to a unique hash. In this manner we can ensure the isomorphic uniqueness of the patterns of graphs. We remove duplicate hashes because we are not considering the graph frequency at this time. The frequency of graph usage may have additional information, but our study focuses on a description of structure and operation, and frequency is not directly relevant to these characteristics. If a graph is present, then it is sufficient to demonstrate the dependency structure of the program. If a graph is present in the program, then it is possible for it to influence the operation regardless of frequency, because there is not a limit to the transitivity within the program. The data dependency graphs can be expanded and mined for additional information. The goal of our work was to provide an increase in resolution while still having efficient and compact representations to be used for other tasks. The partial ordered set of hashes provides an efficient way to compare programs while still providing the granularity needed to analyze the structure of their components. These sets are not computationally intractable, and can be efficiently searched.

Added robustness comes from the analysis of data movement. Each hash is representative of a unique pattern of data movement, and this is describing the operation of the program.  Many studies analyze the program as a whole without segmentation.  Unique hashes of data dependency are not able to be collected without segmenting the program into the contiguous sets of instructions of basic blocks.  Since duplicates are ignored, this reduces the amount of noise present in the program, and allows structural properties to be analyzed. This can provide a description of the components within the program, and how they contribute to the overall class. By taking a bottom up approach to the feature representation, we are able to construct descriptions of program components that have not been seen by the system.  We are able to compare the unknown program to existing samples in terms of their components.  Since this is done at the instruction level, it is not required to show that a correlation between operational semantics holds across abstraction layers.

Our dimensionality has increased from the term frequency representation, which was reduced to 32 feature dimensions through the use of stemming.  But, since the $mov$ instruction makes up a majority of the variance, this is the most important dimension.  The best way to obtain additional resolution within this dimension is not apparent.  There is also not a clear way to demonstrate the correlation of components to their class labels.  Since the term co-occurrence will likely be repeated within the program, it is necessary to further analyze the dependency structure.  The dimensionality of the feature space could potentially be very large, but this can be mitigated by searching within localities.  More targeted searches or classification can be obtained depending on the goals of the studies.

Additionally, we no longer have a positively skewed distribution of terms following a power law distribution.  This represents a reduction in the complexity of the data collected, and enables further statistical analysis to be performed on the datasets.

\section{Conclusion}
In this study we have presented a novel feature representation for malicious programs that can be used for malware classification.  We have shown how to construct the features in a bottom-up approach, and analyzed the overlap of malicious and benign programs in terms of their components.  We have shown that our method of analysis offers an increase in feature resolution that is descriptive of data movement in comparison to $tf-idf$ features.

\subsection{Acknowledgements}
This research was supported in part by Air Force Research Lab grant \#FA8650 to the University of Cincinnati.

Conflicts of Interest: The authors declare no conflicts of interest.

\end{document}